\documentclass[aip,jmp,amsmath,amssymb,graphicx,reprint]{revtex4-1}
\usepackage{gensymb}
\usepackage{booktabs}
\usepackage{epsfig}
\draft 

\begin{document}


\title{Effect of interstitial impurities on the field dependent microwave surface resistance of niobium} 



\author{M. Martinello}
\email[]{mmartine@fnal.gov}
\affiliation{Fermi National Accelerator Laboratory, Batavia, Illinois 60510, USA.}
\affiliation{Department of Physics, Illinois Institute of Technology, Chicago, Illinois 60616, USA.}
\author{A. Grassellino}
\affiliation{Fermi National Accelerator Laboratory, Batavia, Illinois 60510, USA.}
\author{M. Checchin}
\affiliation{Fermi National Accelerator Laboratory, Batavia, Illinois 60510, USA.}
\affiliation{Department of Physics, Illinois Institute of Technology, Chicago, Illinois 60616, USA.}
\author{A. Romanenko}
\affiliation{Fermi National Accelerator Laboratory, Batavia, Illinois 60510, USA.}
\author{O. Melnychuck}
\affiliation{Fermi National Accelerator Laboratory, Batavia, Illinois 60510, USA.}
\author{D. A. Sergatskov}
\affiliation{Fermi National Accelerator Laboratory, Batavia, Illinois 60510, USA.}
\author{S. Posen}
\affiliation{Fermi National Accelerator Laboratory, Batavia, Illinois 60510, USA.}
\author{J. F. Zasadzinski}
\affiliation{Department of Physics, Illinois Institute of Technology, Chicago, Illinois 60616, USA.}


\date{\today}

\begin{abstract}

Previous work has demonstrated that the radio frequency surface resistance of niobium resonators is dramatically reduced when nitrogen impurities are dissolved as interstitial in the material. The origin of this effect is attributed to the lowering of the Mattis and Bardeen surface resistance contribution with increasing accelerating field. Meanwhile, an enhancement of the sensitivity to trapped magnetic field is typically observed for such cavities. In this paper we conduct the first systematic study on these different components contributing to the total surface resistance as a function of different levels of dissolved nitrogen, in comparison with standard surface treatments for niobium resonators. Adding these results together we are able to show for the first time which is the optimum surface treatment that maximizes the Q-factor of superconducting niobium resonators as a function of expected trapped magnetic field in the cavity walls. These results also provide new insights on the physics behind the change in the field dependence of the Mattis and Bardeen surface resistance, and of the trapped magnetic vortex induced losses in superconducting niobium resonators.

\end{abstract}


\maketitle 


Nitrogen-doping is a surface treatment which allows nitrogen atoms to be absorbed as interstitial impurities in the niobium lattice. This treatment has shown a dramatic improvement of superconducting radio-frequency (SRF) properties. In particular, the cavity quality factor, or Q-factor ($Q_{0}$), which is inversely proportional to the power dissipated on the cavity walls, can increas by a factor of three at medium values of accelerating field ($E_{acc}=16 MV/m$) \cite{GrasNdop}. 
The Q-factor is determined by the cavity RF surface resistance $R_{s}$: $Q_{0}=G/R_{s}$, where $ G=270$ $\Omega  $ is the geometrical factor which is independent on material properties.
The RF surface resistance can be decomposed in two contributions, one temperature dependent called BCS surface resistance ($ R_{BCS} $), and one temperature independent called residual resistance ($ R_{res} $).

In this letter we analyze for the first time these two surface resistance contributions for bulk niobium resonators, looking at both the mean free path and the RF field dependencies. The findings here reported allow to understand which is the surface treatment that is capable to maximize the cavity Q-factor for a certain RF field, taking into account the external DC magnetic field trapped during the cooldown through the superconducting (SC) transition.

Q-factor maximization is indeed extremely beneficial in order to decrease the cryogenic cost of continuous wave (CW) accelerators. For this reason the Linear Coherent Light Source (LCLS-II) at SLAC embraced the nitrogen doping technology as treatment for the SRF cavities of the new cryomodules \cite{LiepeLSLC}.

The BCS surface resistance was defined by Mattis and Bardeen \cite{MatBard}. Based on the Bardeen-Cooper-Schrieffer theory of superconductivity  \cite{BCS}  $ R_{BCS} $ decays exponentially with the temperature and depends on several material parameters, such as: London penetration depth $\lambda_{L}$, coherence length $\xi_{0}$, energy gap $ \Delta $, critical temperature $ T_{c} $ and mean free path $l$. It is interesting to note that from the Mattis and Bardeen calculation, $R_{BCS}$ as a function of the mean free path shows a minimum around half of the coherence length \cite{HalSurfRes}.

It is well known that nitrogen doping affects the BCS contribution \cite{GrasNdop}, which in contrary of what happens with standard treatments (120 \degree C bake, BCP or EP), decreases with accelerating field. However, the mechanisms that govern this peculiar behavior are not well understood yet, even though some hypotheses have been theorized \citep{GuruField, Binping}.

This letter adds important insight on the BCS surface resistance field dependence, suggesting that the decreasing of $ R_{BCS} $ may be due to an increasing of the energy gap $ \Delta $ with the RF field.

The introduction  of interstitial impurities and the subsequent change in mean free path affects also the residual resistance. Principal sources of residual losses are: condensed gasses, material inclusions, hydrides and trapped magnetic flux \citep{Padam1}. This last contribution defines the trapped flux sensitivity which in turns depends strongly on the mean free path \cite{MioSRFSensitivity}. Here we show a complete and detailed study which gives the most clear picture of trapped flux dissipation in SRF niobium cavities, from its dependence on the mean free path to its dependence on the RF field. This part of the study is of crucial importance in order to understand both the surface treatment of SRF cavities and the level of magnetic field shielding needed for a specific cryomodule, depending on the application.
%

The amount of trapped flux depends on both the amount of external magnetic field which surrounds the cavity during the SC transition, and on the cooldown details, which affects the magnetic flux trapping efficiency and determine the amount of magnetic field trapped. In particular, fast cooldowns with large thermal gradients along the cavity length help to obtain efficient magnetic flux expulsion, while slow and homogeneous cooling through transition leads to full flux trapping \cite{RomExp1, RomExp2, GonNdopLCLS2, MioHorizCool}.



The cavities analyzed are single cell $1.3$ GHz Tesla-type niobium cavities \cite{Tesla} and after the fabrication about 120 $\mu$m are removed from the inner surface via electro-polishing (EP) and then are UHV baked at $800$ \degree C for three hours. If no further treatment are done such cavities are called EP. Other treatments may be done after this step such as: another run of EP, buffer chemical polishing (BCP), $120$ \degree C baking and N-doping followed by EP removal. Details on the nitrogen-doping process can be found elsewhere \cite{GrasNdop, GrassTalkSRF2015}. The surface treatment of the analyzed cavities are summarized in Table 1 of the supplemental material \cite{Supplemental}. 

A schematic of the instrumentation used to characterize the trapped flux surface resistance may be found in Ref. \onlinecite{MioSRFSensitivity} (Fig. 1). Helmontz coils were adopted to adjust the magnetic field around the cavity. Three or four Bartington single axis fluxgate magnetometers were placed around the cavity equator in order to monitor the external magnetic field during the SC transition. The cavity was also equipped with thermometers in order to monitor the cooldown details.
 
In order to estimate the trapped flux surface resistance, each cavity was measured after two different cooldowns: i) compensating the magnetic field outside the cavity in order to minimize its value during the SC cavity transition, ii) cooling slowly the cavity with about $10-20$ mG of external magnetic field.
After each cooldown, the cavities were tested at the vertical test facility  at Fermilab.

In order to distinguish the effect of trapped flux from other contributions to the residual resistance, we define $ R_{res} $ as sum between two terms: the trapped flux residual resistance, $R_{fl}$, and the ``intrinsic" residual resistance, $R_{0}$, therefore:
\begin{equation}
R_{s}(T,B_{trap})=R_{BCS}(T)+R_{fl}(B_{trap})+R_{0} \, \text{.}
\end{equation}

The surface resistance values are experimentally determined from the Q-factor versus accelerating field measurements as $R_{s}=G/Q$.
Since at very low temperatures $R_{BCS}$ becomes negligible, the  Q-factor is measured at 1.5 K and the residual resistance is calculated as $R_{res}=G/Q(1.5 K)$. If during the cooldown the amount of trapped flux is successfully minimized, then: $R_{fl}\simeq 0$ and $R_{res}\simeq R_{0}$.
In order to obtain $R_{fl}\simeq 0$, the magnetic field outside the cavity was compensated during the cooldown through the SC transition. The average value of magnetic field measured at the cavity equator was always lower than $1$ mG. Alternatively, when possible, the measurement was done after a complete magnetic flux expulsion ($B_{SC}/B_{NC}\sim1.77$ at the equator) \cite{PosenExp}. We have observed that these two methods gave the same results within the measurements uncertainties.
On the other hand, after the cavity trapped some external field: $R_{res}(B_{trap})=R_{fl}(B_{trap})+R_{0}$. 

$R_{BCS}$ and $R_{fl}$ are therefore estimated as follows:
\begin{equation}
R_{BCS}(2 K)=R_{s}(2 K)-R_{0} \, \text{,}
\end{equation}
\begin{equation}
R_{fl}(B_{trap})=R_{res}(B_{trap})-R_{0} \, \text{.}
\end{equation}

$R_{res}(B_{trap})$ was always calculated from the RF measurements after slow cooldowns so that the amount of trapped flux tends to the amount of external field: $ B_{trap}\simeq B_{NC} $.

The trapped flux sensitivity describes the amount of cavity losses per unit of trapped flux and can be defined as:
\begin{equation}
Sensitivity=\dfrac{R_{fl}}{B_{trap}} \, \text{.}
\end{equation}

\begin{figure}[t]
\centering
   \includegraphics[scale=0.38]{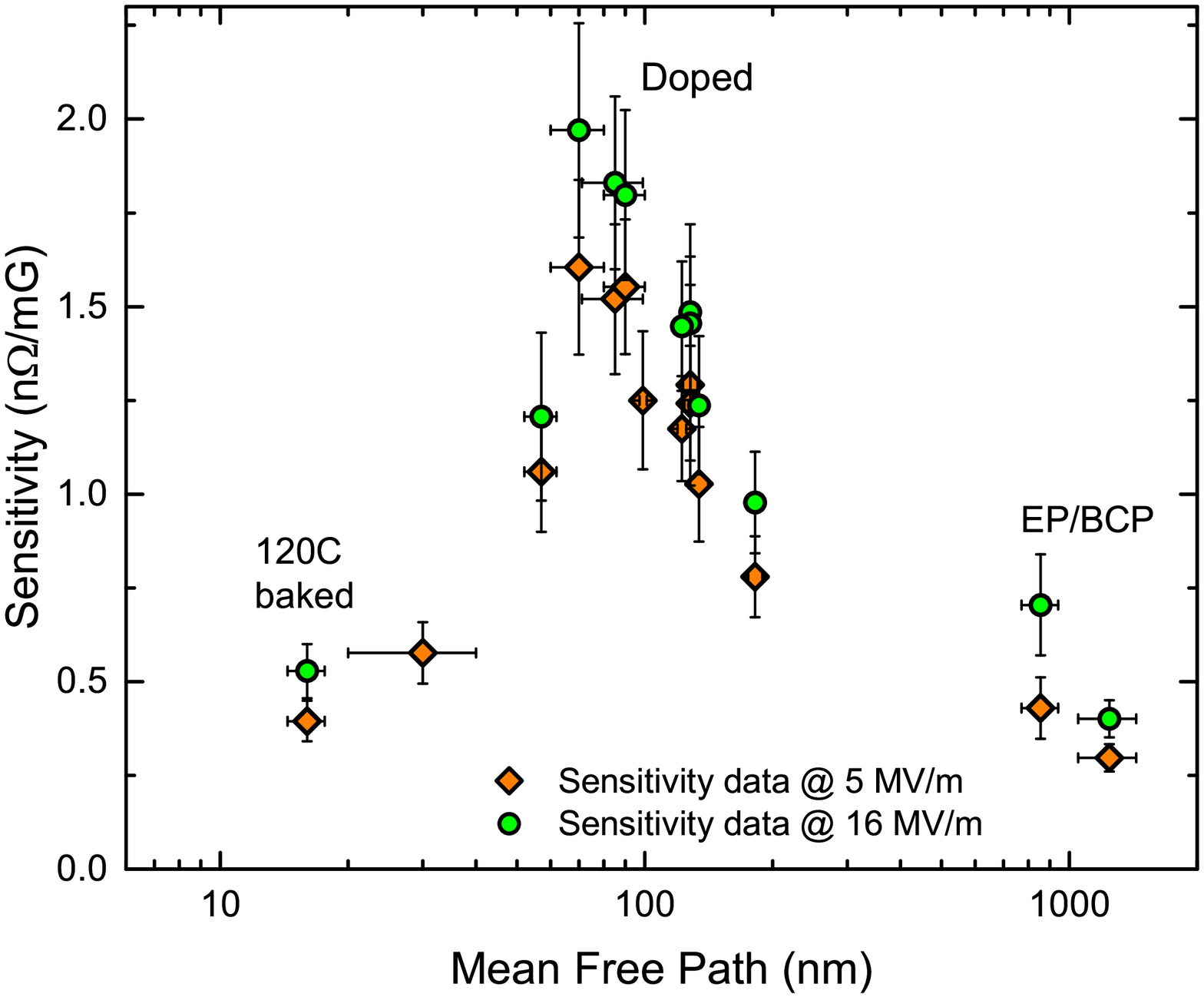}
    \caption{Trapped flux sensitivity calculated at 5 MV/m (orange diamonds) and 16 MV/m (green dots) as a function of the mean free path. The x-axis is break from 160 to 700 nm in order to underline the distribution of doped cavities data points.}
    \label{mfp}
\end{figure}

The values of sensitivity estimated for the cavities analyzed are listed in Table 1 of the supplemental material \cite{Supplemental} and are shown as function of the mean free path in Fig. \ref{mfp}. 

The mean free path of the majority of the cavities analyzed is estimated by means of a $C++$ translated version of SRIMP \cite{HalSrimp} implemented in the OriginLab data analysis program.

The cavity resonance frequency as a function of temperature during the cavity warm up is acquired in order to obtain the variation of the penetration depth, $\Delta \lambda$, as a function of the temperature close to $T_{c}$ \cite{HalSurfRes}. These measurements were done by using a network analyzer which fed the cavity with low power. 

The SRIMP code is used to interpolate $ \Delta \lambda $ versus temperature. The interpolation fixed parameters are: critical temperature \cite{TempCom}, coherence length ($\xi_{0}=38$ nm) and London penetration depth ($\lambda_{L}=39$ nm). The parameters obtained from the interpolation are: mean free path and reduced energy gap (${\Delta}/{kT_{c}}$). 

For the $120$ \degree C bake cavities we used the mean free path measured with LE-$\mu$SR on a representative $120$ \degree C bake cavity cut-out \cite{RomMuSR120C}, since the fit with SRIMP introduces in this case larger error\cite{MfpCom}. 

Figure \ref{mfp} shows that the sensitivity has a bell-shaped trend as a function of the mean free path. The sensitivity is minimized for both very small (120 \degree C bake  cavities) and very large (EP and BCP cavities) mean free paths, and it is maximized around $l\simeq70$ nm. Taking into account N-doped cavities, when over-doped they show the highest sensitivity ($l$ between 70 and 100 nm), while the 2/6 recipe gives the lowest sensitivity ($l$ around $120-180$ nm). From this trend we can also infer that going towards even lighter doping recipes, it should be possible to further decrease the trapped flux sensitivity, without affecting the intrinsic residual resistance.

The experimental data shows some scatter that may be due to uncertainty on the mean free path values, which might be larger than the error bars due to the large number of parameters of the fit. Another reason of this scatter may be the different nature of the cavities analyzed as for example differences in terms of pinning force and dimensions of pinning centers. This parameters become indeed relevant when the dissipation is governed by vortex oscillation \cite{MattiaSensitivityDyn}.

It is well known that the vortex dissipation may be due to two contributions: i) static due to the normal-conducting core of the vortex \cite{Padam1} and ii) dynamic due to the vortex oscillation driven by the Lorentz force in presence of the RF field \cite{GittlemanMixState, Rabinowitz}.  The bell-shaped trend may be found considering simply the static contribution \cite{MattiaSensitivity}, however a better interpolation of the data appears considering dynamic dissipation \cite{MattiaSensitivityDyn}.  The complete set of points published in this letter allowed the completion of an exhaustive model \cite{MattiaSensitivityDyn} able to explain the dependence of the vortex dissipation on the mean free path.

\begin{figure}[t]
\centering
   \includegraphics[scale=0.35]{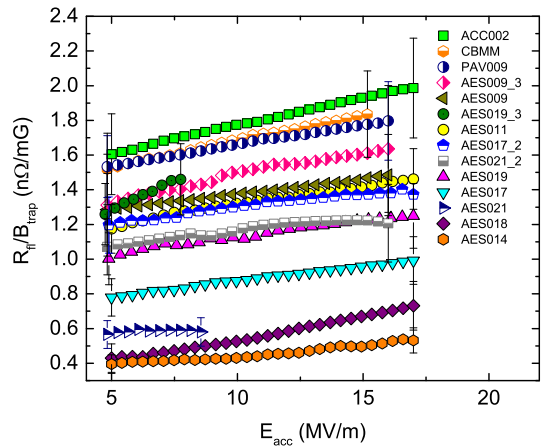}
    \caption{Sensitivity dependence on the accelerating field.}
    \label{SensEacc}
\end{figure}

In Fig. \ref{SensEacc} it can be seen that the trapped flux surface resistance, and therefore the sensitivity, increases with the RF field. A field dependence of $R_{fl}$ was also found studying large grain cavities \cite{GuruDynDiss} and niobium on copper thin film cavities \cite{FilmCERN}. As hypothesized also in Ref. \onlinecite{FilmCERN}, the possible explanation to this phenomenon might be the progressive depinning of vortexes from their pinning center, driven by the increasing of the field.

\begin{figure}[t]
\centering
   \includegraphics[scale=0.35]{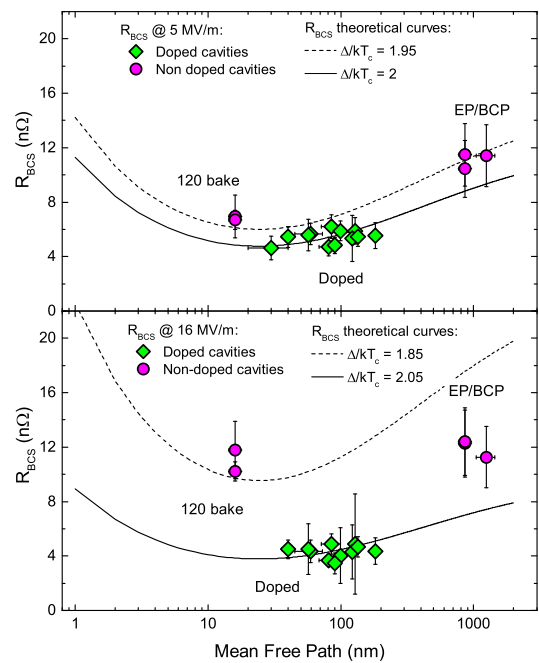}
    \caption{2 K BCS surface resistance as a function of mean free path, at 5 MV/m (upper graph) and 16 MV/m (lower graph). The green diamonds represent doped cavities, while the pink circles represent non doped cavities.}
    \label{BCS}
\end{figure}

The BCS surface resistance at $2$ K is extrapolated after the cooldown with no flux trapped, as the difference between the $R_{s}$ measured at $2$ K and at $1.5$ K (Eq. 2). $R_{BCS}$ measured at low field is shown in Fig. \ref{BCS} as a function of the mean free path. The upper graph shows the results obtained at low field ($5$ MV/m), while the lower at medium field ($16$ MV/m). For EP cavities the mean free path is calculated for one cavity (AES018) and the other one is fixed at the same value, assuming that they should show very similar values. For one N-doped cavity (AES005) the mean free path is directly measured on a cavity cut-out with LE-$\mu$SR \cite{GrassTalkSRF2015}.  

The green diamonds represent the doped cavities, while the pink circles are Niobium cavities with different standard treatments (120 \degree C bake, BCP and EP). The black curves are theoretical curves of $ R_{BCS} $ versus mean free path estimated using SRIMP \cite{HalSrimp} for different reduced energy gap values. 

In both field regimes, doped cavities show lower values of $ R_{BCS} $ than non-doped cavities, proving that $R_{BCS}$ is lowered with the introduction of interstitial impurities. At medium field the difference in $ R_{BCS} $ between doped and non-doped cavities is maximized due to the opposite trend of this surface resistance contribution as a function of the accelerating field.

The values of $ R_{BCS} $ obtained for all the cavities analyzed cannot be interpolated with one single theoretical curve, both at low and medium field, suggesting that the mean free path is not the only parameter changing with the introduction of impurities. Following this hypothesis, one of the other parameters on which the BCS surface resistance depends on ($\lambda_{L}$, $\xi_{0}$, $\Delta$, $T_{c}$) is changing as well. In the low field case, fixing all the other parameters and changing the reduced energy gap $\Delta/kT_{c}$, the 120 \degree C baked, BCP and EP cavities are interpolated with $\Delta/kT_{c}=1.95$, while doped cavities are better interpolated setting $\Delta/kT_{c}=2$. 
At medium field, the difference is even larger being $\Delta/kT_{c}=1.85$ for 120 \degree C baked cavities and $\Delta/kT_{c}=2.05$ for doped cavities. For BCP and EP is probably slightly larger than the value assumed for 120 \degree C baked cavities.

As already mentioned, other parameters may also change with the presence of impurities, however $\Delta/kT_{c}$ seems to be a good candidate since energy gap variations as a function of the impurity content have already been observed in high temperature superconductors \cite{QuazilbashHighT, MiyakawaGap}.

In addition we observe a possible field dependence of the gap. Comparing the upper and lower graph of Fig. \ref{BCS}, for doped cavities $\Delta/kT_{c}$  increases passing from $5$ to $16$ MV/m. This variation may be the reason why $ R_{BCS} $ decreases with the RF field for doped cavities. Increasing of the energy gap with the RF field has been measured in the past \cite{KommersGapEnhance}, and in that case the enhancement of superconductivity was attributed to non-equilibrium effects \cite{EnhanceSupMooij,EliashbergScStimulated}. In the Eliashberg theory the minimum frequency at which non-equilibrium effects may be visible depends on the inelastic collision time of quasi-particles scattering with phonons $ \tau_{E} $ and for niobium this minimum frequency is around 15 GHz \cite{EnhanceSupMooij}. In our case, the introduction of interstitial impurities may cause the increasing of $ \tau_{E} $ allowing non-equilibrium effects to be visible at lower frequencies with respect the case of pure niobium.   

Adding together the measured values of BCS surface resistance and sensitivity, we illustrate for the first time which treatment shows the highest Q-factors depending on the amount of trapped flux.

In order to visualize that, we calculate for each treatments among EP, 120 C baking and N-doping the Q-factor as follows:
\begin{equation}
Q_{0}=G/(R_{BCS}+Sensitivity \cdot B_{trap}+R_{0}) \, \text{,}
\end{equation}
where the values of trapped field, $B_{trap}$, ranges from 0 to 20 mG.

Among the N-doped cavities we chose the 2/6 N-doping treatment, which is the one of greatest interest for high Q application, since it shows a good compromise between $ R_{BCS} $ and sensitivity values exploited so far.

\begin{figure}[t]
\centering
   \includegraphics[scale=0.35]{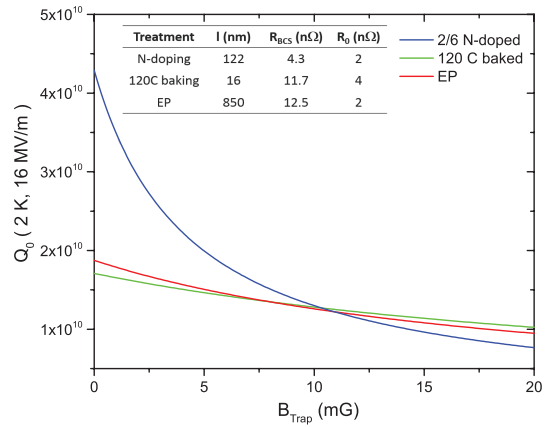}
    \caption{Q-factor at 2 K and 16 MV/m as a function of the trapped field for 120 \degree C bake, EP and 2/6 N-doped cavities (in table are listed the parameters used for the calculation).}
    \label{Qsim}
\end{figure}

The Q-factors as a function of $B_{trap}$ is shown in Fig. \ref{Qsim}. From this graph it is clear that the 2/6 N-doped cavity shows the highest values of Q-factor as long as the trapped field is lower than 10 mG, i.e. within the range of realistic values of magnetic field achievable in modern cryomodules. 

The intrinsic residual resistance depends on many parameters, some related to the surface treatments and others related to the bulk itself \cite{Padam2}. As a matter of fact, 120 \degree C baked cavities usually show value of $R_{0}$ greater than both EP and optimized N-doped cavities \cite{GrasNdop, RomField}. Because of that, the calculation was performed assuming as intrinsic residual resistance: $R_{0}=4$ n$\Omega$ for the 120 \degree C baked cavity and $R_{0}=2$ n$\Omega$ for both EP and N-doped cavities, which are common values found for these treatments. 

Using these findings as guidance, our future work will investigate the performance of N-doped cavities with even larger mean free path values. The treatment that produces best compromise between BCS surface resistance and sensitivity is probably still unexplored.

This letter provides new insight on the physics behind the lowering of the BCS surface resistance, suggesting that with the introduction of impurities another material parameter must change other than the mean free path. The most likely candidate is the energy gap. In addition, the anti Q-slope of N-doped cavities may be explained as the decreasing of the energy gap with the field caused by microwave-driven non-equilibrium effects.

Concluding, from a practical point of view, these results are of crucial importance in order to identify the best surface treatment that allows to reach the highest Q-factors, taking into account all the surface resistance contributions and their dependencies on mean free path, RF field and DC external magnetic field.

This work was supported by the United States Department of Energy, Offices of High Energy and Nuclear Physics and by the DOE HEP Early Career grant of A. Grassellino. Authors would like to acknowledge  technical assistance from the FNAL cavity preparation and cryogenics teams, in particular A. C. Crawford, D. Bice, G. Kirschbaum and D. Marks. Fermilab is operated by Fermi Research Alliance, LLC under Contract No. DE-AC02-07CH11359 with the United States Department of Energy.



%
%

%



\begin{thebibliography}{34}
\bibitem [{Tem()}]{TempCom}%
  \BibitemOpen
  \href@noop {} {}\bibinfo {note} {The critical temperature was estimated for
  each cavity directly from the measurement of the resonance frequency as a
  function of temperature during the warm-up through the normal-conducting
  transition. The resonance frequency drops at transition, and stabilizes when
  the cavity is normal-conducting, allowing the visualization of the critical
  temperature value.}\BibitemShut {Stop}%
\bibitem [{Mfp()}]{MfpCom}%
  \BibitemOpen
  \href@noop {} {}\bibinfo {note} {The $120$ \degree C baking treatment
  modifies the mean free path of the very surface of the cavity, about $60$ nm,
  and for temperatures close to $T_{c}$ the penetration depth becomes larger
  than the modified layer, about thousands of nanometers, probing a region
  which is not representative of the mean free path in the RF layer at low
  temperature.}\BibitemShut {Stop}%
\end{thebibliography}
%
\end{document}